\documentclass{elsart}

\usepackage{graphicx,natbib,amssymb}

\def\lambdabar{\protect\@lambdabar}
\def\@lambdabar{
\relax
\bgroup
\def\@tempa{\hbox{\raise.73\ht0
\hbox to0pt{\kern.25\wd0\vrule width.5\wd0
height.3pt depth.3pt\hss}\box0}}
\mathchoice{\setbox0\hbox{$\displaystyle\lambda$}\@tempa}
{\setbox0\hbox{$\textstyle\lambda$}\@tempa}
{\setbox0\hbox{$\scriptstyle\lambda$}\@tempa}
{\setbox0\hbox{$\scriptscriptstyle\lambda$}\@tempa}
\egroup
}

\begin{document}

\begin{frontmatter}

\title{Cosmological implications of massive gravitons}

\author{Donald H. Eckhardt}
\address{Canterbury, NH 03224-0021, USA}

\author{Jos\'{e} Luis G. Pesta\~{n}a}%\corauthref{cor}}
\address{Departamento de F\'{\i}sica, Universidad de Ja\'en, Campus Las Lagunillas, Ja\'en 23071, Espa\~{n}a}
%\corauth[cor]{Corresponding author.}
%\ead{jlg@ujaen.es}

\and

\author{Ephraim Fischbach}
\address{Department of Physics, Purdue University, West Lafayette, IN 47907-2036, USA}

\begin{abstract}
The van Dam-Veltman-Zakharov (vDVZ) discontinuity requires
that the mass $m$ of the
graviton is exactly zero, otherwise measurements of the
deflection of starlight by the
Sun and the precession of Mercury's perihelion would conflict
with their theoretical values.
This theoretical discontinuity is open to question
for numerous reasons.
In this paper we show from a phenomenological viewpoint that the $m>0$ hypothesis is in accord with Supernova Ia
and CMB observations,
 and that the large scale structure of the universe
suggests that $m \sim 10^{-30}~$eV$/c^2$.
\end{abstract}


\end{frontmatter}

\section{The vDVZ discontinuity}

Recent papers have revisited the question of
whether gravity could be mediated by an ultralight but
massive graviton \citep{Will:06,CNPT:05,DR:05,GN:09,AW:09}.
The classic works of \citet{vDV:70} and \citet{Z:70} demonstrated that,
in a linearized theory of massive gravity,
the tensor gravitons of General
Relativity (GR), which couple to $T_{jk}$, the stress-energy tensor, 
are supplemented by a scalar graviton, which couples to $T_{jj}$, the trace of the stress-energy tensor.  In the open limit as $m\rightarrow 0+$, the gravitational forces between non-relativistic masses can be accomodated simply by increasing the gravitational constant $G$ by the factor $4/3$ \citep{Ma:08,Zee}; whereas for $m=0$, there is no scalar graviton hence $G$ is not increased to compensate for the discontinuity. An increased $G$ for $m>0$ would  lead to a prediction for
the angle starlight is deflected by the Sun that is $4/3$ that of GR.  The agreement between GR and this key observational test is
now at the $\sim 10^{-5}$ level \citep{Will:06}, which implies that the graviton must be strictly massless.

Various elaborate explanations for resolving the vDVZ discontinuity have been suggested (e.g., \cite{V:72,DDGV,G:05,GG:05}).
In the present paper we follow the approach of \citet{GN:09} and \citet{AW:09}
by sidestepping the theoretical question of whether and how a theory of massive gravitons can be formulated, and focusing instead on observational data. Specifically we show that Supernova Ia and CMB data support the $m > 0$
hypothesis, and that the large scale structure of the universe
suggests that $m \sim 10^{-30}~$eV$/c^2$.

\section{A cosmological expansion discontinuity}

Consider a spatially uniform density ($\rho=\rho(t)$) expanding  universe
for which the spherical coordinate line element is of the form
\begin{equation}
ds^2=c^2\,dt^2-a^2(t)\,[dr^2+f^2(r)\,d\Omega],
\label{eq:q1}
\end{equation}
From the null
geodesic ($ds=0$) for light traveling along a radial path ($d\Omega=0$),
 we then have
\begin{equation}
c\,dt=a\,dr.
\label{eq:dra}
\end{equation}
The Einstein-de Sitter~\citep{P:93}  universe (GR with cosmological constant $\Lambda=0$, $f(r)=r$, $m=0$) requires that
$a\propto t^{2/3}$, 
so if a 
photon is emitted at $t_e$ when $a(t_e)=a_e$ and
observed at $t_o$ when $a(t_o)=a_o$,
the distance between the emitter (e) and observer (o) at time $t_o$
is~\citep{P:93}
\begin{equation}
a_or=(2c/H_o)[1-(1+z)^{-1/2}],
\label{eq:gmassless}
\end{equation}
where $H_o=H(t_o)$ is Hubble's constant, and $z=a_o/a_e-1$ is the
observed redshift.
If, however, $m>0$, then Birkhoff's theorem
\citep{P:93} does not apply, so we cannot model the gravitational field
on the surface of a sphere without considering its surroundings,
even if (as in this model) the surroundings have a constant density.
Instead, we adopt the phenomenological approach of \cite{Ma:08} who
 hypothesized
 that the gravitational potential energy of two point masses, $m_1$ and $m_2$,  separated by the distance $r$ is
\begin{equation}
V_Y(r)=-\frac{Gm_1m_2}{r}\;\exp(-\mu r),
\end{equation} 
where $\mu=mc/\hbar$.  Maggiore, following the \cite{V:72} scenario, used $G=(4/3)G_N$, where $G_N$ is the Newtonian gravitational constant, but we shall ignore any distinction (if indeed one exists) between $G$ and $G_N$.   Then for a uniform density $\rho$ universe, the gravitational potential at $r=0$ is formally
\begin{equation}
\Psi_Y=-4\pi G\rho\int_0^{\infty}\frac{\exp(-\mu r)}{r} \, r^2\,
dr=-4\pi G\rho\lambdabar^2=-3GM_\lambdabar/\lambdabar,
\label{eq:yint}
\end{equation}
where  $\lambdabar=1/\mu$ is the reduced Compton wavelength of the graviton, and $M_\lambdabar$ is the mass of a sphere with radius $\lambdabar$.
(This  outwardly naive solution does not apply if $m=0$. See Appendix~\ref{ap:NY} for its rigorous derivation and a discussion of the $m=0$ discontinuity.)
The integral of Eq.~\ref{eq:yint} is independent of where we select the origin, so the potential  has no gradient and there is no cosmological acceleration; that is, $\dot{a}=\mathrm{constant}$ and
Eq.~\ref{eq:gmassless} is not valid.
On a cosmological scale this is a Milne universe, for which the Einstein tensor $G_{ik}$ must be zero.  To effect this,  we set
\begin{equation}
a(t)=a(0)+\dot{a}t,
\label{eq:q2}
\end{equation}
and, in Eq.~\ref{eq:q1},
\begin{equation}
%f(r)=(c/\dot{a})\sinh(\dot{a}r/c)
f(r)=\frac{c}{\dot{a}}\sinh\frac{\dot{a}r}{c}.
\label{eq:q3}
\end{equation}
One can easily verify that Eqs.~\ref{eq:q1}, \ref{eq:q2} and \ref{eq:q3} define a metric for which $G_{ik}=0$  by using Parker's Mathematica~\citep{math} notebook, {\it Curvature and the Einstein Equation}~\citep{Ha:02}.  Appendix~\ref{ap:met} details how Eq.~\ref{eq:q3} was derived and verified.

For $m>0$, the integral of Eq.~\ref{eq:dra} is
\begin{equation}
r=\int_{t_e}^{t_o}\frac{c\,dt}{a}=\frac{c}{\dot{a}}\int_{a_e}^{a_o}\frac{da}{a}=\frac{c}{\dot{a}}\ln(a_o/a_e)
=\frac{c}{\dot{a}}\ln(1+z).
\label{eq:rz}
\end{equation}
On
substituting $\dot{a}=a_oH_o$,
 we find that the distance between the emitter and observer at $t_o$ is
\begin{equation}
a_or=(c/H_o)\ln(1+z).
\label{eq:gmassive}
\end{equation}
The area of the wavefront at $t_o$ is $4\pi(a_of)^2$, where
\begin{equation}
a_of = \frac{c}{H_o}\sinh\left[\ln(1+z)\right] =\frac{c}{2H_o}\left[1+z-\frac{1}{1+z}\right],
\label{eq:area2}
\end{equation}which, except for the sign, is in agreement with \citet{SDJ:05} for a linear coasting cosmology (their Eq.~8, with $z_1\rightarrow z$, $z_2\rightarrow 0$).

\section{The agreement with the type Ia supernovae (SNe Ia) observations}

Two large research teams \citep{Retal:98,Petal:99} recently
found discrepancies between the distances to high
redshift ($z\sim 1$) Type Ia supernovae (considered to be well
understood ``standard candles") when the SNe Ia distances are
determined by their apparent magnitudes versus when they are determined
by their redshifts using Eq.~\ref{eq:gmassless}.
The SNe Ia
brightnesses appear to be about 25\% weaker than expected, and
so their distances are correspondingly greater than their redshifts
would indicate.  To phrase it another way, the SNe Ia
redshifts are smaller than their magnitudes would indicate, and
so $\dot{a}$ in the past
appears to be smaller than it is today; that is, $\ddot{a}> 0$ and the
expansion rate of the  universe is accelerating.
Actually, however, the acceleration satisfies
$\ddot{a}>\ddot{a}_{\mathrm{model}}\propto
-t^{-4/3}$ if the model is that of an Einstein-de Sitter universe.
Although $\ddot{a}$ is not necessarily greater than zero at
present, there still is an unmodeled effect, an ostensibly repulsive
force that  has been ascribed to
``dark energy", that enters the GR field
equations as a non-zero $\Lambda$.
We offer
an alternative to dark energy, namely that gravitons are not massless.

\begin{figure*}
\begin{center}
\includegraphics[width=\columnwidth]{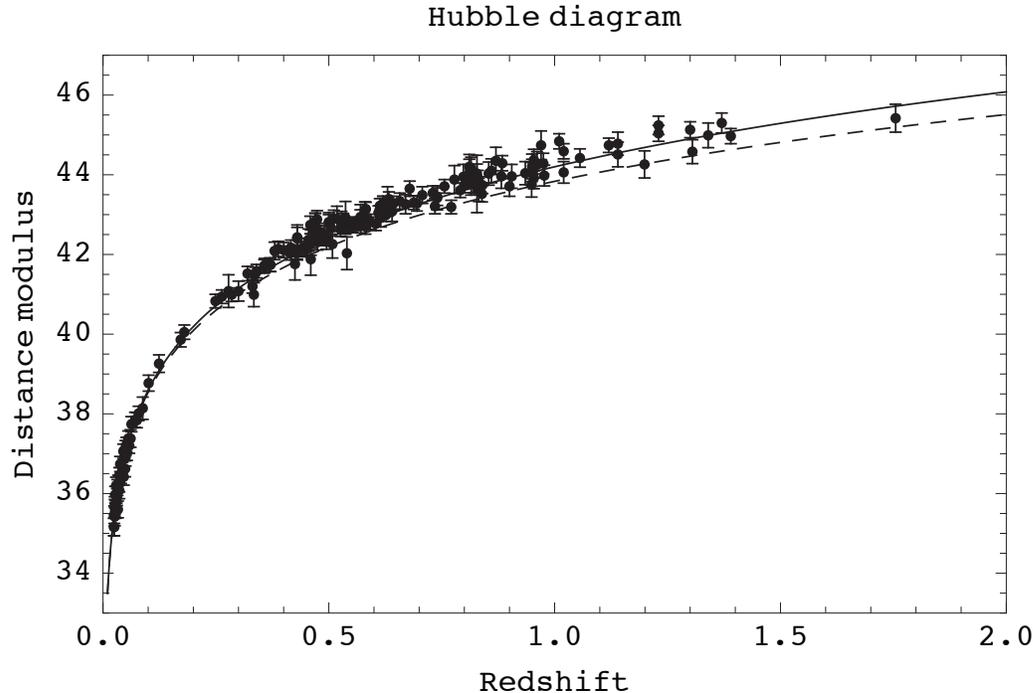}
\caption{Hubble diagram fits for 182 high-confidence SNe Ia from
\citet{Retal:07}.
The dashed line is a plot of Eq.~\ref{eq:gmassless}  (flat universe with $m=0$),
and the solid line is a plot of Eq.~\ref{eq:gmassive} ($m>0$).  The horizontal
axis is the redshift $z$, and the vertical axis is the photometrically determined distance modulus.
The Hubble constant for each curve has been chosen to give the best fit to the observations.
}
\label{fig:sne}
\end{center}
\end{figure*}

In Fig.~\ref{fig:sne} we compare
the most recent set of high-confidence (``gold'') \citep{Retal:07}
data for 182 SNe Ia
sources distributed over the interval $0.0233 \lesssim z \lesssim 1.755$
with two models in
which $\Lambda=0$: GR for a flat  universe, with $m=0$ (Eq.~\ref
{eq:gmassless}),
and GR with $m>0$ (Eq.~\ref{eq:gmassive}).
The Hubble constant
was determined by a weighted least-squares
adjustment of the model to the data.
For $m > 0$, $H_o = 60.1~ \mathrm{km\;s}^{-1} \mathrm{\, Mpc}^{-1}
(\chi^2 = 179)$, and for
$m=0$, $H_o = 55.4~ \mathrm{km\;s}^{-1} \mathrm{\, Mpc}^{-1}(\chi^2 =  288)$.
Because the number of degrees
of freedom $\nu = 180$ is large, $x = (\chi^2 - \nu)/\sqrt{2\nu}$
is approximately
normally distributed with unit variance~\citep{C:58, AS:64}.
For $m > 0$, $x = -0.1$, whereas for $m=0$, $x = 5.9$.
The $m>0$ model 
is quite plausible, but the
$m=0$ model
is not; and the estimated $H_o$ for the $m>0$
model is appreciably closer to that of \cite{Retal:98} than is the
 estimate for
$m=0$.
Indeed, the $m>0$ model's Hubble constant is even closer to the $H_o=62.3~ \mathrm{km\;s}^{-1} \mathrm{\, Mpc}^{-1}$ estimate of \cite{TSR:08}.

The Einstein-de Sitter ($m=0$) and Milne ($m>0$) models have power-law cosmologies in which the cosmological scale factor evolves as $a\propto t^q$.   Assuming 
a power law parameterization, \citet{SDJ:05} found the exponent that gives
 the best fit
to the first 157 ``gold'' subset of SNe Ia data~\citep{Retal:04} is
 $q=1.04_{-0.06}^{+0.07}$. This provides additional support in favor of Milne's model ($q=1$) over that of Einstein and de Sitter ($q=2/3$), but there are other cosmologies to consider, notably those involving dark matter and dark energy.
 
 \begin{table}[h]
 \label{t1}
\caption{\label{table:res}Qualities of fit between the 182 ``gold'' supernovae for three cosmologies. All the calculations have been independently done for this paper. 
Where $\lambdabar\ll c/H_o=L_H$ (the Hubble length), the $m>0$ model is equivalent to a Milne universe.  The conventional cosmological parameters, $\Omega_m,\,\Omega_\Lambda,\,\mathrm{and}\;\Omega_R$, are  associated with the density, the cosmological constant, and the curvature; and   
%=H_o/(100\,\mathrm{km\;s}^{-1} \mathrm{\, Mpc}^{-1})$
$h$  is the Hubble parameter.}
\begin{tabular}{lcccccr}
\hline\hline
Universe&$\Omega_m$&$\Omega_{\Lambda}$&$\Omega_{R}$&$h$&$\chi^2$&$x$\\
\hline
$\Lambda$CDM&0.29&0.71&0.00& 0.647& 164&-0.8\\
Einstein-de Sitter&1.00&0.00&0.00& 0.554& 288&+5.7\\
Milne&0.00&0.00&1.00&0.601&179&-0.1 \\
\hline\hline
\vspace{1mm}
\end{tabular}
\end{table}
 
 \citet{Retal:07} proffered the $\Lambda$CDM model with cold dark matter and dark energy ($\Omega_m=0.29, \Omega_\Lambda=0.71$), fitting their ``gold'' data set with $\chi^2=150$, which is appreciably less than our Milne fit's $\chi^2=179$.  However, their Milne fit's $\chi^2=164$ is also less than ours; yet they fit the Einstein-de Sitter model, with $\chi^2=285$, which is close to our $\chi^2=288$.  Our calculations differed somewhat from theirs, and we could discern no pattern to the offsets, so we recalculated their $\Lambda$CDM model fit, and found $\chi^2=164$, and $x=-0.8$.
 Our fits to the $\Lambda$CDM, Einstein-de Sitter, and Milne models are summarized in 
 Table 1.  The Hubble parameter, $h=H_o/(100\,\mathrm{km\;s}^{-1} \mathrm{\, Mpc}^{-1})$,  is the only free parameter for each fit.  The results are independent of the two fitting technique we used - marginalization and conventional weighted least-squares.  Our $h$ estimates are valid only if there are no systematic uncertainties, including that of the SN Ia absolute magnitude. As an example of the sensitivity, a systematic uncertainty of $\pm 0.03$ mag \citep{Hetal:09} would cause $h$ to be offset by $\mp 0.0083$.  Even so, the systematic uncertainties affect only the $h$ estimates, but not the $\chi^2$ results.

   If the criterion for the better fit is having the lower $\chi^2$, then the $\Lambda$CDM model ($\chi^2=164$) clearly edges the Milne model  ($\chi^2=179$).  However, if we give full credence to the observation variances (that is, if we assume that they are not exaggerated), then the criterion for the better fit is having the higher likelihood.  The maximum likelihood is at the peak of the $x$ normal distribution, $x=0$, so then the Milne model ($x=-0.1$) edges the $\Lambda$CDM model ($x=-0.8$), but this comparison is for solutions that minimize $\chi^2$ rather than minimize $x$.   In both cases $x<0$, so by varying the solution parameters from the least-squares determinations, $x$ will increase from its minimum, and there will be non-unique solutions for which $x=0$.  For example, the only parameter for the $m>0$ model is $h$, and the two roots to $x(h)=0$ are $h=0.596$ and $h=0.607$.  With three independent parameters ($h,\,\Omega_m,\,\Omega_\Lambda$), the $\Lambda$CDM model is more complicated because it involves a hypersurface rather than a curve.  By varying each parameter, one at a time, from the $\chi^2$ minimum, six solutions for $x=0$ can (in principle) be found; but if they are all allowed to vary together, then there will be a continuum of solutions. 
The relative plausibility of these two models cannot be definitively settled by statistics.  We prefer instead to invoke Occam's razor: (1) the $\Lambda$CDM model requires cosmological attractions by vast amounts of dark matter that are largely countered by cosmological repulsions by vast amounts of dark energy, whereas (2) the $m>0$ model merely requires that the graviton not be massless.  Occam's razor favors the second.     
  Moreover, by not requiring a $\Lambda$ term, we avoid conceptual difficulties, such as those described by
\citet{C:01}, who considers a non-zero $\Lambda$ to be exceedingly problematic,
but who also concedes that denying its existence in face of GR and existing data is even more problematic.
The massive graviton
 alternative has the additional advantage that it is not
sensitive to the graviton mass, provided only that $m$ is not identically zero.  Finally,
if there are no massless gravitons,  then there is no critical density below
which the  universe is open; the  universe is necessarily open.

\section{The graviton mass}
\label{gm}

Although the results in Fig. 1 are insensitive to the specific 
value of $m$ provided that $m\not= 0$, it is interesting to consider the limits
on $m$ that do exist.
Following \citet{Will:06}
we assume that in the absence of other long
range forces, the effect of a massive graviton is to replace the
Newtonian gravitational potential $-GM/r$ of a point mass $M$
by the Yukawa potential $-(GM/r)e^{-r/\lambdabar}$.
Hence
any system of characteristic size $R$ whose behavior
is correctly described by Newtonian gravity implies a limit
$\lambdabar \gtrsim  R$.  Several authors \citep{Will:06,Will:98,GN:74,V:98}
have analyzed data for various astrophysical systems to set limits on
$\lambdabar$.  Using solar system data from \citet{TBHS:88},
Will finds $\lambdabar > 2.8 \times 10^{12}$ km, and from
galaxy supercluster  data one can infer $\lambdabar \gtrsim
6 \times 10^{19}$ km \citep{Will:06,Will:98,GN:74,V:98}.
Other lower limits on $\lambdabar$ are summarized in \citet{Will:06}.
Utilizing the previous arguments we can infer an upper limit for $\lambdabar$.
Suppose that there is an immense volume of uniform density matter whose
extent is many skin depths ($\lambdabar$'s).  On a point mass that is several skin
depths inside the outer surface of the volume, the specific force due to the
Yukawa potential is zero, since the source appears the same in all 
directions.
Suppose next that this volume is divided into two parts: one is a
sphere of radius $r \ll$ $\lambdabar$, and the other part is the remainder of the
volume with the sphere removed.  The sphere is the source of a force
directed toward its center (which can easily be quantified), so the remainder
of the volume is  the source of a force of an identical magnitude, but
directed away from the center of  the sphere.
Hence if there exists a vast uniform density volume, and somewhere 
deep inside it there is a density perturbation that makes the local density smaller than the 
mean density, then the lower density matter will be accelerated outward.  At the outer boundary of 
the volume matter will, of course, be accelerated inwards.  The volume will transform toward 
being a hollow bubble until the thickness of the higher density bubble matter is 
just a few skin depths.
There are some additional considerations: conservation of angular  momentum places a
constraint on how far the inward moving matter can progress although 
it does not, by itself, constrain how far the outward moving matter can go.
According to \citet{DSZ:90},
``clusters of galaxies form surfaces with a thickness of about 10-20 Mpc
surrounding empty regions with characteristic size of 100-200 Mpc." 
With the estimate 
\begin{equation}
\label{eq:lest}
\lambdabar \approx 5\,\mathrm{Mpc},
\end{equation}
 the gravitational force is
significant for galaxy superclusters and also between neighboring
superclusters that form a ``surface,"  but negligible in the
vast empty regions between the surfaces.
A skin depth $\lambdabar = 5$ Mpc $=1.6 \times 10^{20}$ km
corresponds to $m=1.3\times 10^{-30}$~eV$/c^2$; hence we
conjecture that $m\sim 10^{-30}$ eV$/c^2$.
It is then conceivable that there could be heavier gravitons in addition to this lightest graviton, with masses somewhere in the range 
\begin{equation}
1.3\times 10^{-30} \mbox{ eV}/c^2 <m< 7.4\times 10^{-23} \mbox{ eV}/c^2,
\label{range}
\end{equation} 
which corresponds to
\begin{equation}
1.6\times 10^{20}\mathrm{\, km}> \mathrm{\lambdabar} >2.8\times 10^{12}\mathrm{\, km}.
\end{equation}
The Laser Interferometer Space Antenna 
(LISA) gravitational wave detector would not be able to detect the $m\sim 10^{-30}$ eV$/c^2$ mass, but it might be sensitive to the more massive gravitons in the range of Eq.~\ref{range} if, indeed, any exist~\citep{Will:98}.

\section{The cosmic microwave background (CMB)}
If $m>0$, the dynamics of a baryon fluid at the epoch of recombination differ substantially from those of conventional models which tacitly assume that the graviton is massless.  In Section~\ref{gm}, we explained why a massive body with dimensions that are large compared with $\lambdabar$ would not be stable, but instead would tend to separate into smaller bodies with dimensions that are of the order of a few skin depths or less - that is, to the the scale of galaxy superclusters.  We hypothesize that this instability is the principal cause of CMB power spectrum anisotropies.  On recombination, massive bodies with breadths larger than, say, $D=4\lambdabar$ (that is, with 
radii exceeding two skin depths if the bodies are spheres) would divide into lesser sized bodies separated from each other by the distance of approximately $D$.  Because the universe has expanded, the nominal separation distance of observed radiation peaks from the last scattering surface, with redshift $z_\star$, is $(1+z_\star)D$; and the nominal angular separation of peaks on the celestial sphere for $z_\star^2\gg1$ is (see Eq.~\ref{eq:area2}) 
\begin{equation}
\gamma=\frac{(1+z_\star)D}{a_of}\approx \frac{2H_oD}{c}=\frac{8\lambdabar}{L_H},
\label{eq:gam}
\end{equation} 
where $L_H=c/H_o$ is the Hubble length.  Setting $h=0.60$ (from Table~\ref{t1}) and $\lambdabar\approx 5$ Mpc (Eq.~\ref{eq:lest}), we calculate $\gamma\approx  0.008\,\mathrm{radians}\approx 0.5^\circ$,  which is close to what has been observed \citep{Letal:01,Setal:03}.
%\footnote{We arrived at this estimate without fudging.  Considering the fuzziness of our guesstimates, inserting the fudge factor $\pi/4$ now would do no harm, yet it would simplify  Eq.~\ref{eq:gam} to $\gamma\approx\lambda/L_H$, where $\lambda$ is the (unreduced) Compton wavelength.}

\section{Summary}

Massive gravitons lead to an
alternative relation between the luminosity distance [$a_or$ versus $a_o f_o$ from Eq.~\ref{eq:area2}]
and the redshift $z$ [Eq.~\ref{eq:gmassive} versus Eq.~\ref{eq:gmassless}].  A comparison of
Eqs.~\ref{eq:gmassive} and~\ref{eq:gmassless} in the framework of a cosmology with $\Lambda=0$
favors the massive graviton alternative, as discussed in the text
and shown in Fig.~\ref{fig:sne}. 
% entail the high cost of extra ``baggage'' (Occam entities).
 Moreover, our graviton mass estimate, $m\sim 10^{-30}$ eV$/c^2$, is consistent with the structure of galaxy superclusters and with the pattern of the cosmic microwave background. Although the redshift
data could be attributable to a $\Lambda$CDM model \citep{Retal:04}, doing so would
conflict with the rule of thumb that ``less is better'' because of the need for additional suppositions.

\section{Conclusions}
The inverse-square law is the \emph{only} possible law that leads to a cosmological contracting acceleration.  Its GR heritage, common acceptance, usage and applicability over broad but limited domain lead us to consider it to be the norm.  It need not be; indeed we might even consider it to be a peculiar exception.  

Any force law that assigns the graviton a mass, $m> 0$, is a local law because the force falls off both geometrically (like the inverse-square law) and exponentially with distance.  In that case, the cosmological principle that ``everything is the same everywhere'' is not even needed.  A weaker principle would suffice, for example the cosmological legal principle that ``every physical law is the same everywhere.'' Gravity pulls the clusters in a supercluster together, but it does not pull distinct superclusters together. The gravity-free Milne model is underlain by local gravity effects that lead to our interesting universe rather than a featureless cloud of expanding matter.  Dark matter and dark energy are then regarded as gratuitous Procrustean artifacts.

\appendix

\section{Contrasting the Newtonian and Yukawa Accelerations in a Uniform Density Universe}
\label{ap:NY}
We contrast the familiar Newtonian acceleration with the less familiar Yukawa acceleration, starting with  a model  for which
  there is only a thin spherical shell of mass $dM$ at $r=a$.

\subsection{Newtonian Potential Inside the Shell}
The Newtonian potential $d\Psi_N(r,a)$ at the origin is
$d\Psi_N(0,a)=-G\,dM/a$, and the Newtonian potential satisfies the homogeneous equation
$\nabla^2d\Psi_N=0$ (in free space), whose solution is a linear combination of $d\Psi_N=\mathrm{constant}$ and $d\Psi_N\propto 1/r$.  The only such solution that satisfies $d\Psi_N(0,a)=-G\,dM/a$ is 
\begin{equation}
d\Psi_N(r,a)=-G\,dM/a,\;r\leq a.
\label{eq:ni}
\end{equation}
The solution interval is closed at the shell because the potential (but not its gradient) is continuous there.
The potential has no gradient for $r<a$, so there is no Newtonian gravitational acceleration inside the shell.  This is Newton's ``iron sphere theorem''~\citep{P:93}.

\subsection{Yukawa Potential Inside the Shell}
The Yukawa potential at the origin is
$d\Psi_Y(0,a)=-G\,dM \exp(-\mu a)/a$.
  The Yukawa potential in free space satisfies the homogeneous equation $\nabla^2d\Psi_Y-\mu^2d\Psi_Y=0$, whose solutions are linear combinations of $\exp(\pm\mu r)/r$.  The only such combination that equals $d\Psi_Y(0,a)$ at the origin is $d\Psi_Y(0,a) \sinh(\mu r)/(\mu r)$, so the Yukawa potential inside and on the shell is
\begin{equation}
d\Psi_Y(r,a)=-G\,dM\,\frac{\exp(-\mu a)}{a}\;\frac{\sinh(\mu r)}{\mu r}\equiv \frac{\partial w_1(r,a)}{\partial a}\,da,\; r\leq a.
\label{eq:yi}
\end{equation}

\subsection{Newtonian Potential Outside the Shell}
The solution to $\nabla^2d\Psi_N=0$ that equals $d\Psi_N(a,a)$ at $r=a$ and approaches zero as $r\rightarrow\infty$ is
\begin{equation}
d\Psi_N(r,a)=-G\,dM/r,\;r\geq a.
\label{eq:no}
\end{equation}
Outside the shell, the Newtonian gravitational acceleration (the gradient of the potential), follows the inverse square law.

\subsection{Yukawa Potential Outside the Shell}
The solution to $\nabla^2d\Psi_Y-\mu^2d\Psi_Y=0$ that equals $f_1(a,a)$ at $r=a$ and remains finite outside the shell is
\begin{equation}
d\Psi_Y(r,a)=-G\,dM\,\frac{\sinh(\mu a)}{\mu a}\;\frac{\exp(-\mu r)}{r}\equiv \frac{\partial w_2(r,a)}{\partial a}\,da,\; r\geq a.
\label{eq:yo}
\end{equation}

Note that $\partial w_1(r,a)/\partial a=\partial w_2(a,r)/\partial a$.

We next integrate over $a$ to derive the Newtonian and Yukawa gravitational potentials in a uniform density universe, for which $dM=4\pi\rho a^2\,da$.  For $\Psi_N$, this is a trivial exercise.The integral of
Eq~.~\ref{eq:no} is $-GM/r\propto - r^2$, where $M$ is the mass of the sphere $r\le a$; this is the familiar Newtonian gravitational potential. The integral of
 Eq.~\ref{eq:ni} is $-G(M_U-M)/a$, where $M_U$ is the mass of the universe inside the cosmic horizon.   The radial gradient of the Newtonian potential is $GM/r^2$; at $r=a$, it proportional to $a$, so there is a gravitational attraction between any two points that is proportional to their separation distance.  If $m=0$, the universe has a uniform contracting gravitational acceleration.
 
\subsection{Derivation of Eq.~\ref{eq:yint}}

The integral of Eq.~\ref{eq:yi} is
\begin{equation}
W_1(r)=\int_r^\infty \frac{\partial w_1(r,a)}{\partial a}\,da=-4\pi G\rho\lambdabar^2
\,\frac{\exp(-\mu r)(1+\mu r)\sinh(\mu r)}{\mu r},
\end{equation}
and the integral of Eq.~\ref{eq:yo} is
 \begin{equation}
W_2(r)=\int_0^r \frac{\partial w_2(r,a)}{\partial a}\,da=-4 \pi G\rho\lambdabar^2
\,\frac{\exp(-\mu r)[\mu r\cosh(\mu r)-\sinh(\mu r)]}{\mu r}.
\end{equation}  
Their sum is
$W_1(r)+W_2(r)=-4\pi G\rho\lambdabar^2=-3GM_{\lambdabar}/\lambdabar$, the same as Eq.~\ref{eq:yint}.
If $m>0$, the universe has no gravitational acceleration.

If we let $\mu\rightarrow 0$ in Eqs.~\ref{eq:yi} and \ref{eq:yo}, then they become Eqs.~\ref{eq:ni} and \ref{eq:no}; there is no $m=0$ discontinuity for thin shells.  However, if we first integrate over all shells, then $\Psi_Y$ is independent of $r$, whereas $\Psi_N\propto 1/r$; letting $\mu\rightarrow 0$ after the integration does not change this situation.

\section{Derivation of the  Metric}
\label{ap:met}

We used Leonard Parker's Mathematica~\citep{math} notebook, {\it Curvature and the Einstein Equation}~\citep{Ha:02}, to determine $f(r)$ in the Eq.~\ref{eq:q1} metric, when the distance formula is Eq.~\ref{eq:q2}.  The only change we made to the notebook was in the definition of the metric, using these two statements:

a=a0+adot t

metric=\{\{-a$^\wedge$2,0,0,0\},\{0,-(a f[r])$^\wedge$2,0,0\},\{0,0,-(a f[r])$^\wedge$2 Sin[$\theta$]$^\wedge$2,0\},\{0,0,0,c$^\wedge$2\}\}. 

Then the $G_{22}$ ($G_{\theta\theta}$) component of the Einstein tensor turned out to be
\begin{equation}
G_{22}=f(r)\left[f''(r)-(\dot{a}/c)^2\right].
\end{equation}
Eq.~\ref{eq:q3} is an evident solution to $G_{22}=0$, so we substituted it back into the Mathematica notebook by defining

f[r]=(c/adot) Sinh[adot r/c],

and all components of the Einstein tensor turned out to be zero.  QED.   

\section{Acknowledgements}

We thank the referee for several helpful comments.

The work of EF was supported in part by the U.S. Department of Energy under Contract No. DE-AC02-76ER071428.


\begin{thebibliography}{}

\bibitem[{{Abramowitz} \& {Stegun}(1964)}]{AS:64}
{Abramowitz}, M., \& {Stegun}, I. A., 1964, Handbook of Mathematical Functions (Washington, DC: National Bureau of Standards)

\bibitem[{{Arun} \& {Will}(2009)}]{AW:09}
{Arun}, K. G., \& {Will}, C. M. 2009, Class. Quantum Grav. 26 (2009) 155002

\bibitem[{{Carroll}(2001)}]{C:01}
{Carroll}, S. M. 2001, Living Rev. Relativity 4, 1

\bibitem[{{Cram\'{e}r}(1958)}]{C:58}
{Cram\'{e}r}, H. 1958, Mathematical Methods of Statistics (Princeton: Princeton University Press)

\bibitem[{{Creminelli} {et~al.}(2005)}]{CNPT:05}
{Creminelli}, P., {Nicolis}, A., {Papucci}, M., \& {Trincherini}, E. 2005, J. High Energy Phys. 9, 3

\bibitem[{{Deffayet} {et~al.}(2002)}]{DDGV}
{Deffayet}, C., {Dvali}, G., {Gabadadze}, G., \& {Vainstein}, A. 2002, Phys.Rev. D 65, 044026

\bibitem[{{Deffayet} \& {Rombouts}(2005)}]{DR:05}
{Deffayet}, C., \& {Rombouts}, J. W. 2005, Phys. Rev. D 72, 044003

\bibitem[{{Dolgov}, {Sazhin} \& {Zeldovich}(1990)}]{DSZ:90}
{Dolgov}, A. D., {Sazhin}, M. V., \& {Zeldovich}, Ya. B. 1990, Basics of Modern Cosmology (Gif-sur-Yvette: Editions Fronti\`{e}res)

%\bibitem[{{Elgar{\o}y} \& {Multam\"{a}ki}(2007)}]{EM:07}
%{Elgar{\o}y}, O., \& {Multam\"{a}ki}, T. 2007, A\&A 471, 65

\bibitem[{{Gabadadze} \& {Gruzinov}(2005)}]{GG:05}
{Gabadadze}, G., \& {Gruzinov}, A. 2005,  Phys. Rev. D 72, 124007 

%\bibitem[{{Galli} {et~al.}(2008)}]{GBMS:08}
%{Galli}, S., {Bean}, R., {Melchiorri}, A., \& {Silk}, J. 2008, Phys. Rev. D 78, 3532

\bibitem[{{Goldhaber} \& {Nieto}(2009)}]{GN:09}
{Goldhaber}, A. S., \& {Nieto}, M. M. 2009, Rev. Mod. Phys. in press

\bibitem[{{Goldhaber} \& {Nieto}(1974)}]{GN:74}
{Goldhaber}, A. S., \& {Nieto}, M. M. 1974, Phys. Rev. D 9, 1119

\bibitem[{{Gruzinov}(2005)}]{G:05}
{Gruzinov}, A. 2005, New Astron. 10, 311

\bibitem[{{Hartle}(2002)}]{Ha:02}
{Hartle}, J. B. 2002, Gravity: An Introduction to Einstein's General Relativity (Boston: Addison-Wesley Pub Co.)

\bibitem[{{Hicken} {et~al.}(2009)}]{Hetal:09}
{Hicken}, M., {et~al.} 2009, ApJ accepted

%\bibitem[{{Komatsu} {et~al.}(2009)}]{Ketal:09}
%{Komatsu}, E., {et~al.} 2009, ApJS 180, 330

\bibitem[{{Lange} {et~al.}(2001)}]{Letal:01}
{Lange}, A. E., {et~al.} 2001, Phys.Rev. D 63, 042001

\bibitem[{{Maggiore}(2008)}]{Ma:08}
{Maggiore}, M. 2008, Gravitational Waves (Oxford: Oxford University Press)

%\bibitem[{{Wang} \& {Mukherjee}(2006)}]{WM:06}
%{Wang}, Y., \& {Mukherjee}, P. 2006, ApJ 650, 1

\bibitem[{{Peebles}(1993)}]{P:93}
{Peebles}, P. J. E. 1993, Principles of Physical Cosmology (Princeton: Princeton University Press)

\bibitem[{{Perlmutter} {et~al.}(1999)}]{Petal:99}
{Perlmutter}, S., {et~al.} 1999, ApJ 517, 565

\bibitem[{{Riess} {et~al.}(1998)}]{Retal:98}
{Riess}, A. G., {et~al.} 1998, AJ 116, 1009

\bibitem[{{Riess} {et~al.}(2004)}]{Retal:04}
{Riess}, A. G., {et~al.} 2004, ApJ 607, 665

\bibitem[{{Riess} {et~al.}(2007)}]{Retal:07}
{Riess}, A. G., {et~al.} 2007, ApJ 659, 98

\bibitem[{{Sethi}, {Dev} \& {Jain}(2005)}]{SDJ:05}
{Sethi}, G., {Dev}, A., \& {Jain}, D. 2005, Phys. Lett. B 624, 135

\bibitem[{{Spergel} {et~al.}(2003)}]{Setal:03}
{Spergel}, D. N., {et~al.} 2003, ApJS 148, 175

\bibitem[{{Talmadge} {et~al.}(1988)}]{TBHS:88}
{Talmadge}, C., {Berthias}, J.-P., {Hellings}, R. W., \& {Standish}, E. M. 1988, Phys. Rev. Lett. 61, 1159

\bibitem[{{Tammann}, {Sandage} \& {Reindl}(2008)}]{TSR:08}
{Tammann}, G. A., {Sandage}, A., \& {Reindl}, B. 2008, AAR 15, 289

\bibitem[{{Vainshtein}(1972)}]{V:72}
{Vainshtein}, A. I. 1972, Phys. Lett. B 39, 393

\bibitem[{{van Dam} \& {Veltman}(1970)}]{vDV:70}
{van Dam}, H., \& {Veltman}, M. 1970, Nucl.~Phys.~B 22, 397

\bibitem[{{Visser}(1998)}]{V:98}
{Visser}, M. 1998, Gen. Relativ. Gravit. 30, 1717

\bibitem[{{Will}(1998)}]{Will:98}
{Will}, C. M. 1998, Phys. Rev. D 57, 2061

\bibitem[Will(2001)]{Will:06}
Will, C. M. 2001, Living Rev. Relativity 4, 4

\bibitem[{{Wolfram}(1999)}]{math}
{Wolfram}, S. 1999, The Mathematica Book (Wolfram Media/Cambridge University Press)

\bibitem[{{Zakharov}(1970)}]{Z:70}
{Zakharov}, V. I. 1970, JETP Lett. 12, 312

\bibitem[{{Zee}(2003)}]{Zee}
{Zee}, A. 2003, Quantum Field Theory in a Nutshell (Princeton University Press)

\end{thebibliography}
\end{document}